\documentclass[proof]{WileyASNA-v1}

\articletype{Article Type}%

\received{}
\revised{}
\accepted{}
\newcommand{\Msun}{$M_\odot$} 
\newcommand{\farcs}{\mbox{$.\!\!^{\prime\prime}$}}
\newcommand{\degs}{\mbox{$^{\circ}$}}
\usepackage{graphicx}
\usepackage{hyperref}
\begin{document}

\title{Relativistic X-ray jets at high redshift}

\author[1]{Daniel A Schwartz*}
\author[1]{Aneta Siemiginowska}

\author[2]{Diana M Worrall}

\author[2]{Mark Birkinshaw}

\author[3]{Teddy Cheung}

\author[4]{Herman Marshall}

\author[5]{Guilia Migliori}

\author[6]{John Wardle}

\author[7]{Doug Gobeille}

\authormark{Daniel A Schwartz \textsc{et al}}

\address[1]{\orgdiv{High Energy Astrophysics}, \orgname{Smithsonian
    Astrophysical Observatory}, \orgaddress{\state{Massachusetts}, \country{USA}}}

\address[2]{\orgdiv{HH Wills Physics Laboratory}, \orgname{University of Bristol}, \country{UK}}

\address[3]{\orgdiv{Space Science Division}, \orgname{Naval Research
    Laboratory}, \orgaddress{\state{Washington D C}, \country{USA}}}

\address[4]{\orgdiv{Kavli Institute}, \orgname{MIT}, \orgaddress{\state{Massachusetts}, \country{USA}}}

\address[5]{\orgname{INAF Istituto di Radioastronomia},  \country{Italy}}

\address[6]{\orgdiv{Physics Department}, \orgname{Brandeis University}, \orgaddress{\state{Massachusetts}, \country{USA}}}

\address[7]{\orgdiv{Physics Department}, \orgname{University of Rhode Island},
  \orgaddress{\state{Rhode Island}, \country{USA}}}

\corres{*60 Garden St., Cambridge, MA 02138, USA. \email{dschwartz@cfa.harvard.edu}}

\abstract{Powerful radio sources and quasars emit relativistic jets of
  plasma and magnetic fields that travel hundreds of kilo-parsecs,
  ultimately depositing energy into the intra- or inter-cluster
  medium. In the rest frame of the jet, the energy density of the
  cosmic microwave background is enhanced by the bulk Lorentz factor
  as $\Gamma^2$, and when this exceeds the magnetic energy density the
  primary loss mechanism of the relativistic electrons is via inverse
  Compton scattering. The microwave energy density is also enhanced by
  a factor (1+z)$^4$, which becomes important at large redshifts. We
  are using \textit{Chandra} to survey a z>3 sub-sample of radio
  sources selected with 21 cm-wavelength flux density > 70 mJy, and
  with a spectroscopic redshift. Out of the first 12 objects observed,
  there are two clear cases of the X-rays extending beyond the
  detectable radio jet.}


\jnlcitation{\cname{%
\author{Schwartz, D. A.  \textsc{et al}}} (\cyear{2019}), 
\ctitle{Relativistic X-ray jets at high redshift},
\cjournal{Astronomische Nachrichten}, \cvol{2019;}.}

\fundingInfo{}

\maketitle

\section{Introduction}

The number of quasars known in the redshift range z$>$3 and beyond z=6
has exploded in recent years, with the Sloan Digital Sky Survey (SDSS)
playing a key role. At redshift 7 the Universe is less than a billion
years old, and at redshift 3 only about 2 billion years, or 15\% of
the current age estimate based on the Planck satellite measurements
\citep{Planck17} of the cosmic microwave background (CMB)\footnote{In
  this article we will use the parameters H$_0$=67.3 km s$^{-1}$
  Mpc$^{-1}$, $\Omega_m$=0.315, and $\Omega$=1 from those Planck
  results}.  Following these results, we expect the facilities coming
on line in the next decade, such as SKA, the LSST, JWST, and the
proposed  \textit{Lynx} X-ray observatory, to allow us to trace formation and
growth from the first massive black holes and galaxies to the present
cosmic epoch.  X-ray observations will play a crucial role in
delineating this development of structure in the Universe. The
proposed  \textit{Lynx} observatory (https://arxiv.org/abs/1809.09642), with an
effective area of 2 m$^2$ and angular resolution of 0\farcs5 will have
the capability to detect black holes of only 10$^4$ \Msun at redshift
z=10 in the deepest surveys. However, even prior to  \textit{Lynx}, all-sky
X-ray surveys, and high resolution observations with \textit{Chandra},
may reveal previously unknown active galaxies at redshifts z>5 via
their X-ray jet emission \citep{Schwartz02}.

A \textit{Chandra} survey of radio quasars with arcsec scale radio
jets detected about 2/3 of the targets in short, 5 to 10 ks,
observations \citep{Marshall18}. The redshifts of those quasars ranged
from 0.1 to 2.1. The X-ray emission is interpreted most simply as
radiation from inverse Compton up-scattering of the cosmic microwave
background photons (IC/CMB) within a kpc-scale jet that is moving
relativistically (bulk Lorentz factor $\Gamma \ge $3) with respect to
the co-moving frame of the parent quasar. The bulk relativistic motion
explains the one-sided nature of the jets via Doppler boosting, and is
indicated by apparent superluminal motion and by radio brightness
temperatures in excess of 10$^{12}$ K on pc and sub-pc scales.  In the
presence of magnetic fields and photons, relativistic electrons will
scatter off both, emitting synchrotron radiation and IC radiation. The
dominant energy loss for the electrons simply depends on the energy
density of the target fields or photons. The Compton scattering is
often called ``external Compton'' to distinguish from synchrotron
self-Compton emission.  In the rest frame of a relativistic jet, the
external photons will appear highly anisotropic, and their energy
density appear enhanced by a factor approximately $\Gamma^2$. The
scattered radiation is highly anisotropic in the frame of an observer
that is nearly at rest with respect to the CMB. The intensity is
enhanced, or diminished, by a factor $\delta^{3+\alpha}$ where
$\delta=1/(\Gamma(1-\beta \cos[\theta]))$ is the Doppler factor. Here
$\alpha$ is the spectral index of the radiation (defined for flux
density $\propto \nu^{-\alpha}$), $\Gamma=1/\sqrt(1-\beta^2)$ is the
bulk Lorentz factor of the jet and $\theta$ is the angle from the jet
direction to the observer line of sight.

\begin{figure}[h]
\includegraphics[width=.95\columnwidth]{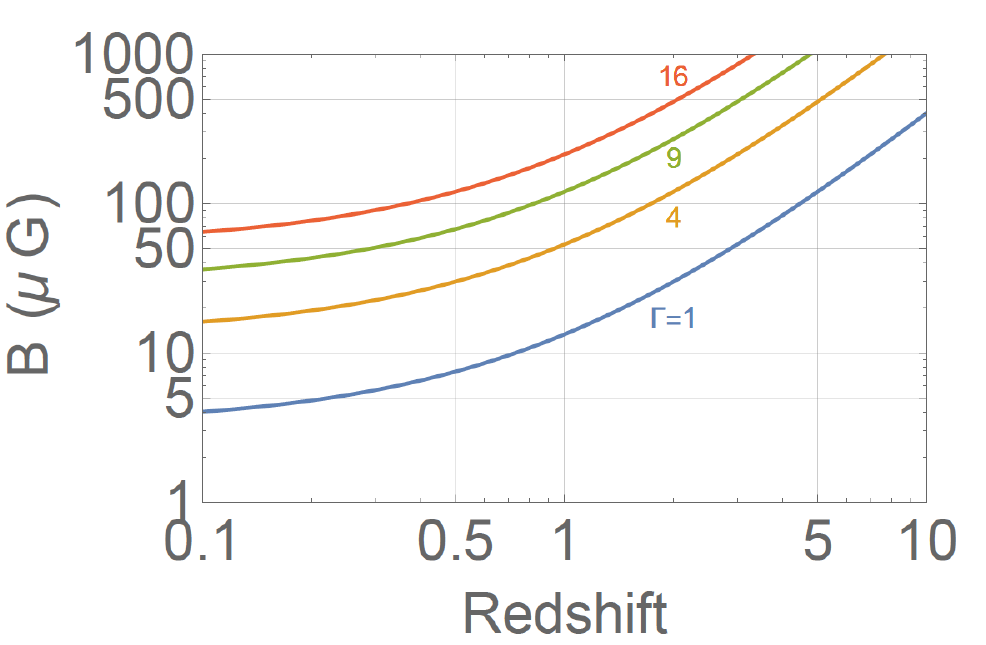}
\caption{ Criteria for IC/CMB radiation to dominate electron energy loss. The line for each value of
  $\Gamma$  plots the magnetic field strength for which the magnetic energy density equals the CMB
  energy density as a function of redshift. For a jet at a given
  redshift, the IC/CMB mechanism will dominate unless the magnetic
  field is stronger than that given on the curve of the 
appropriate Lorentz factor  $\Gamma$.
\label{fig:bEQCMB}}
\end{figure}

At lower redshifts the IC/CMB mechanism is not usually dominant.
X-rays from FR I type radio jets are best explained as an extension of
the radio synchrotron spectrum \citep{Harris06}. For several quasars
the gamma-rays which are predicted by IC/CMB fall above upper limits
from the Fermi gamma-ray telescope \citep{Breiding17}. However, since
the energy density of the CMB increases with redshift as (1+z)$^4$,
IC/CMB eventually becomes dominant. Figure~\ref{fig:bEQCMB} plots the
magnetic field strength vs. redshift at which magnetic and CMB energy
densities become equal. This occurs at redshifts \citep{Schwartz02}
\begin{equation}
\label{eq:CMBdominates}
z\geq\max{[(0.556\sqrt{B_{\mu G}/\Gamma} -1),0]}.
\end{equation}
The figure shows four cases, $\Gamma$ = 1, 4, 9, and 16.
For magnetic field strengths below the curve for the given value of
$\Gamma$, IC/CMB will dominate. At redshifts greater than 3 this
occurs for magnetic fields weaker than 50 $\mu$Gauss, even for bulk
motion with $\Gamma \approx 1$, and occurs for fields weaker than 100's of $\mu$Gauss
for modest values of $\Gamma$. For comparison, magnetic fields in
lower redshift jets are typically estimated to be a few 10's of
$\mu$Gauss.

Very near a quasar, photons from the accretion disk, the broad line
clouds, and the dusty torus will dominate.  The IC/CMB will be the
dominant external Compton mechanism at distances from the quasar core
\begin{equation}
\label{eq:CMBdominatesvsExternal}
r_{kpc}\geq 14.13\sqrt L/(\Gamma (1+z)^2), 
\end{equation}
where L is the radiative luminosity of the quasar in units of
10$^{44}$ erg s$^{-1}$.

\begin{figure}[h]
\includegraphics[width=.95\columnwidth]{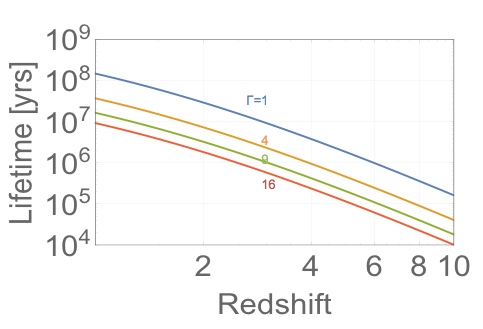}
\caption{Lifetime of electrons emitting approximately 1 keV X-rays via
  the IC/CMB mechanism. From top to bottom, the curves give the
  lifetime for X-ray jets with bulk Lorentz factors $\Gamma$= 1, 4, 9, and
  16, respectively. Losses due to synchrotron radiation are assumed to
  be much less than to IC/CMB.\label{fig:lifetime}}
\end{figure}

Jets at high redshift should increasingly manifest as X-ray jets
rather than radio jets. Two factors favor the X-ray emission. The
first results from the cosmological diminution of surface brightness,
which is proportional to (1+z)$^{-4}$. This reduces the ability to
detect extended radio sources. However, for IC/CMB X-ray emission this
is compensated by the (1+z)$^4$ increase of the CMB energy
density. The second factor is due to the shorter lifetimes of the
electrons emitting GHz synchrotron radiation. 
Very roughly, electrons with Lorentz factor 1000/$\Gamma$ are
producing the 1 keV radiation.  Figure~\ref{fig:lifetime} shows the
lifetime for such electrons to lose energy via IC/CMB as a function of
redshift and of the jet bulk Lorentz factor. GHz synchrotron radiation in
a 10 $\mu$Gauss field is produced  by electrons with $\gamma
\approx$10$^4$. Such electrons will up-scatter the CMB to energies of
100~$\Gamma$  keV,
while also emitting GHz synchrotron radiation. So at redshift 3, for a
typical $\Gamma$ of 10, the X-ray jet lifetime is 100 times longer
than the GHz radio jet.

The radio source J0730+4049 at z=2.50, (B3 0727+409), was
discovered with a 12\farcs\  long X-ray
jet with no extended radio emission except for one knot 1\farcs4 along
the X-ray jet from the quasar core
\citep{Simionescu16}. Based on that discovery, and the
rationale discussed above, we have undertaken a \textit{Chandra} survey for
more systems dominated by X-ray jets.

\section{The high redshift survey}

Our sample is based on a survey conducted by
\cite{Gobeille14}. That survey covered the area jointly
observed  in the FIRST radio survey \citep{Becker95} and
optically by the Sloan Digital Sky Survey (SDSS). It contains all 123
quasars at redshifts $\ge$2.5 with a spectroscopic redshift measured
in the SDSS, and with a radio flux density greater than 70 mJy at
1.4 GHz \citep{Gregory96}.  The radio flux was summed from the total
system, including any extended emission. Of the quasars, 61 showed
extended emission. Thirty of these were classified as triples, and as
such we considered that they were probably not beamed in our
direction. Of the remaining 31, we selected the 16 with redshift z$>$3
as most likely to be detectable IC/CMB radiators in 10 ks  \textit{Chandra}
observations. The two at the largest redshifts, J1430+4204 at z=4.7
and J1510+5702 at z=4.3, had previously been detected at 1.9 counts
ks$^{-1}$ \citep{Cheung12} and 1.4 counts ks$^{-1}$
\citep{Siemiginowska03}, respectively, and were not
re-observed.

%\begin{center}
\begin{table}[h]%
%\centering
\caption{\textit{Chandra} High Redshift Sample: z$>$3\label{tab1}}
\tabcolsep=0pt%
\begin{tabular*}{20pc}{@{\extracolsep\fill}lccl@{\extracolsep\fill}}
\toprule
& &{\textbf{Predicted }}& {\textbf{Detected}}
 \\
& & {\textbf{Counts}}& \textbf{Jet}\\
\textbf{Source Name} & \textbf{Redshift}  & \textbf{per 10 ks}    & \textbf{Counts?}   \\
\midrule
J1435+5435  & 3.809   & 9.5   & NO     \\
J0833+0959  & 3.731   & 8.9   & ?   \\
J1223+5038  & 3.501   & 7.3   & NO     \\
J0933+2845  & 3.421   & 6.8   & NO      \\
J0909+0354  & 3.288   & 6.0   & NO    \\
J0801+4725  & 3.267   & 5.9   & NO    \\ 
J1655+1948  & 3.26    & 5.8   & NO     \\
J1616+0459  & 3.215   & 5.6   & NO     \\
J1405+0415  & 3.209   & 5.6   & YES, 7    \\
J1655+3242  & 3.189   & 5.5   & NO     \\
J1610+1811  & 3.118   & 5.1   & YES, 8    \\
J1016+2037  & 3.115   & 5.1   & NO     \\
J1128+2326  & 3.049   & 4.8   & ?   \\
J0805+6144  & 3.033   & 4.7   & NO    \\
\bottomrule
J1430+4204  & 4.72    &19  & YES     \\
J1510+5702  & 4.30     &14       & YES   \\
\bottomrule
\end{tabular*}
\begin{tablenotes}%%
\item J0833+0959 and J1128+2326 have not yet been observed.
%\item[1] \cite{}
%\item[2] Example for a second table footnote.
\end{tablenotes}
\end{table}
%\end{center}

\section{Detection of X-ray jets}

Table~\ref{tab1} shows the sample of 14 objects, plus the two
previously detected. The 14 were accepted as 10 ks \textit{Chandra}
observations in cycle 19. The first two columns give an abbreviated source name,
and the redshift.  In column 3 we predicted the counts expected in the
0.5--7 keV band if
their flux scaled from that of J1430+4204 and J1510+5702 simply by the
ratio of (1+z)$^4$ for their redshifts. The counting rates of those
two sources previously detected do have a ratio very nearly
(5.72/5.30)$^4$. We expected this number to be a conservative mean,
with some jets much brighter than predicted. The last column indicates
whether an X-ray jet was detected. For J1405 and J1610 we do have
statistically significant detections, with the number of counts as
tabulated.  For 10 other quasars, we do not have significant indication of a
jet.

For the systems without an apparent jet, the exact upper limit must be
carefully considered for each case. To get a rough idea, since we
typically can expect one scattered X-ray from the quasar core, and
negligible background in 10 ks, four
counts would be a detection to 98\% confidence for any one
source. However, since we have 14 targets, we require 5 counts such
that none of the 14 would give a spurious detection to 98\% confidence.

When we do detect N $<$5 counts, then we must ask how many counts might
have been expected such that to 98\% confidence only N resulted from
our observation. If N is 4, the upper limit would be 10.6. That
calculation assumes that we have one specific area to search, e.g.,
along the direction of a radio jet, knot, or lobe. In view of these
statistical complications, we must perform further analysis  
for the sources which are not detected, and  only enter a ``NO'' in Table~\ref{tab1}. We present preliminary results on the
two detections.

\subsection{J1405+0415}

\begin{figure}[t]
\includegraphics[width=.95\columnwidth]{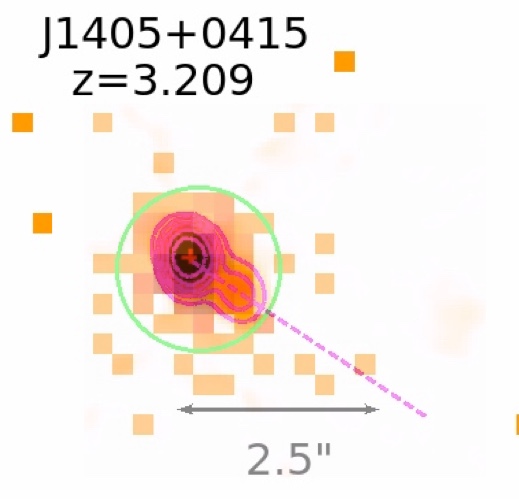}
\caption{Contours of the 6.2 GHz radio emission, superposed on the
  pixelated X-ray counts. Radio contours (magenta in the on-line
  version) start at 3.3 mJy/beam and are logarithmically spaced to the
  peak of 694 mJy/beam. Restoring beam is 0\farcs39 $\times$
  0\farcs32. X-ray pixels are 0\farcs24 square. Quasar peak is 31
  counts per pixel. Outside the 1 arcsec radius circle centered on the
  quasar (green in the on-line version) the pixels have at most one
  count. The dashed line extension from the quasar through the radio
  knot suggests the direction of a jet, and 7 X-ray photons are found
 along this line where only 0.9 are expected from background and from
 scattered quasar photons. \label{fig:J1405}}
\end{figure}

Figure~\ref{fig:J1405} shows the 0.5 to 7 keV X-ray data from our
observation of J1405+0415. We have superposed contours from our JVLA
observations centered at 6.2 GHz. The X-ray jet is seen as an
extension along the line at position angle 231\degs\ from the quasar
core through the radio knot 0\farcs8 away. There is no radio emission
detected along this line past 0\farcs8. The knot has a flux density 28
mJy and a radio spectral index $\alpha$=0.91$\pm$0.09.  At position
angle 255\degs\ there is a very faint radio lobe 3\farcs3 from the
core, with flux density 2.9
mJy and spectral index 1.66$\pm$0.4. 

In the core, \cite{Yang08} measure many components with VLBI and VSOP
data. The 4.86 GHz VSOP data show a range of position angles from
282\degs\ to 335\degs\, at distances 0\farcs5 to 13\farcs2 from the
core, and with sizes 0\farcs3 to 2\farcs5. From several components in
the core which show brightness temperatures in excess of 10$^{12}$ K
they conclude there is relativistic motion.

The jet X-ray flux is 9$\times$10$^{-15}$ erg cm$^{-2}$ s$^{-1}$, which
would correspond to a luminosity of 8$\times$10$^{44}$ erg s$^{-1}$ if
it were isotropically emitted. The quasar core has 269 photons, giving
an X-ray flux of 3.3$\times$10$^{-13}$  erg cm$^{-2}$ s$^{-1}$ and a
luminosity of 3.2$\times$10$^{46}$ erg s$^{-1}$. Thus the jet X-ray
flux is 2.6\% that of the quasar core, with a factor of 2
uncertainty.

\subsection{J1610+1811}

\begin{figure}[t]
 \includegraphics[width=.95\columnwidth]{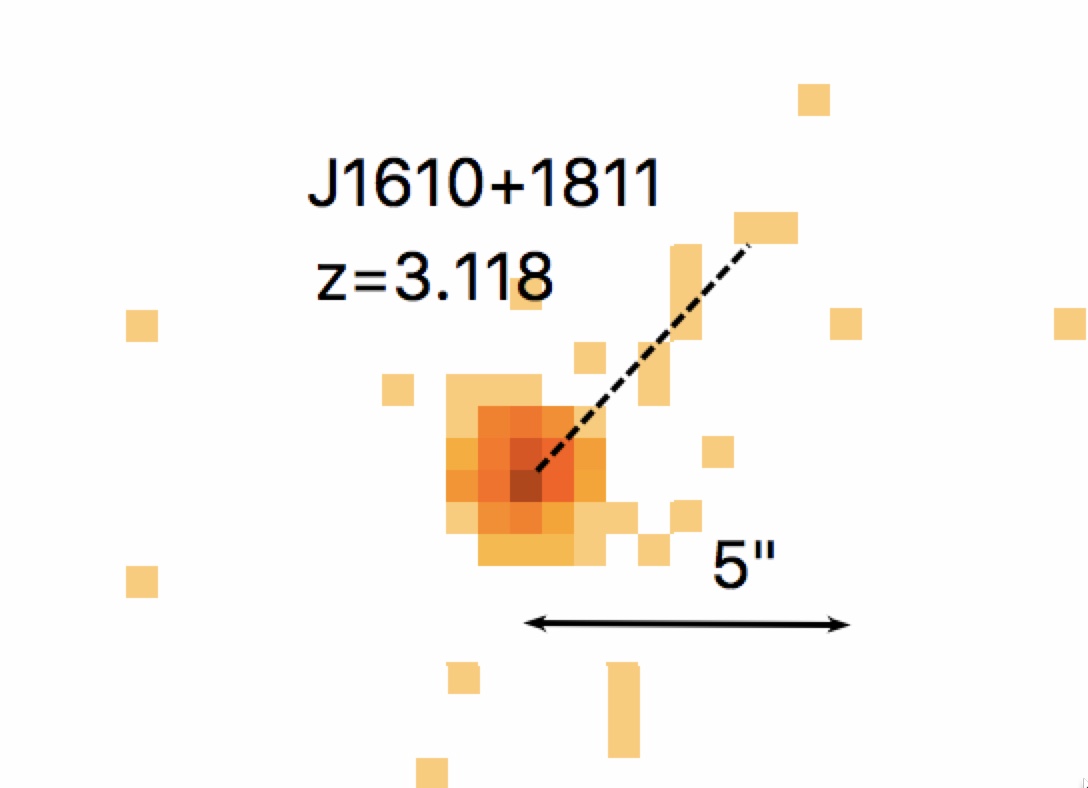}
 \caption{The 0.5--7 keV X-ray counts are binned in 0\farcs49 pixels.
   The quasar peak has 104 counts. More than 2 pixels (1 arcsec) from
   the core, pixels have at most one count. The dashed line at
   position angle 316.5\degs\ is on the line from the quasar to an
   extended radio lobe 4\farcs8 away. The eight X-ray counts which are
   along this line constitute a significant detection of a jet. No
   radio emission is detected between the quasar and the lobe in our
   JVLA observations centered at 6.2 GHz.\label{fig:J1610}}
\end{figure}

Figure~\ref{fig:J1610} plots the 0.5 to 7 keV X-ray data from our
observation of J1610+1811. There is significant X-ray emission along
the dashed line that extends from the 
quasar radio core to an extended radio lobe. The 8 counts associated
with the jet correspond to a flux of 10$^{-14}$ erg cm$^{-2}$
s$^{-1}$. If interpreted as isotropic emission, the luminosity would
be 9$\times$10$^{44}$ erg s$^{-1}$. Those numbers could be a factor of
2 larger or smaller due to the detection being just above threshold.
The quasar core has 331 X-ray counts. This is a measured flux of
5.2$\times$10$^{-13}$ erg cm$^{-2}$ s$^{-1}$ or an isotropic luminosity of
4.6$\times$10$^{46}$ erg s$^{-1}$. The X-ray jet flux is apparently
2\%  (between1\% and 4\% to 95\% confidence) of the quasar X-ray flux.

A 2 mas long jet extends in the same direction in an 8.4 GHz VLBI
observation by \cite{Bourda11}. Our JVLA observations centered at 6.2
GHz show no indication of emission outside of the 0\farcs5 region
around the quasar, up to the edge of the lobe at 4\farcs4 from the
core, with a 1 mJy upper limit. At 6.2 GHz the flux density is 65 mJy
from the core, and 8.2 mJy from the NW lobe.

The radio lobe and the VLBI structure each
 show a one-sided system, which indicates a relativistic jet.
We can construct an illustrative IC/CMB model by taking the radio flux
density to equal 1 mJy at 6.2 GHz. We use usual assumptions, (cf.
\cite{Schwartz06,Worrall06}) of minimum energy in the magnetic field
and relativistic particles, relativistic electrons with a power law
spectral index 2.1 from $\gamma$=30 to $\gamma$=10$^6$, a jet width of
2 kpc, charge balance provided by equal numbers of protons, and
uniformly tangled and isotropic distribution of particles and fields
in the jet rest frame.  We make the assumption that the bulk Lorentz
factor equals the Doppler factor. This is equivalent to assuming the
jet is at the largest possible angle to the line of sight for the
given Doppler factor. We find $\delta$=$\Gamma$=9.4, a magnetic
field strength 12 $\mu$Gauss, and a kinetic power 1.4$\times$10$^{46}$
erg s$^{-1}$. Such a jet would be at an angle 6.2\degs\ to our line
of sight.

Electrons emitting the 6.2 GHz radiation have Lorentz factors
$\gamma\approx$ 11,000. Their lifetime against energy loss by IC
scattering of the CMB is only 8500 years, compared to 880,000 years
for the electrons with $\gamma \approx$ 100 which are emitting 1 keV
X-rays. The short radio lifetime shows that it is reasonable that only
the X-rays are detectable.

With the true radio flux density being less than 1 mJy, both the
magnetic field strength and the relativistic particle density will be
smaller if we preserve the minimum energy assumption. In that case the
jet must have a larger bulk Lorentz factor since the CMB energy
density in the jet must be enhanced further in order to produce the
same X-ray emission. Conversely, if the X-ray flux is at the lower
limit allowed, the magnetic field strength and particle densities
would be larger, and the jet Lorentz factor smaller. 

\section{Discussion}

The numbers of predicted counts, column 3 in Table~\ref{tab1}, assumes
that the jets are identical in their intrinsic properties, magnetic
field, particle density and spectrum, bulk Lorentz factor, age and
size, and furthermore at the same angle to our line of sight as the
object J1430+4204 from which they are scaled. This is clearly
unrealistic, but does correctly predict for at least 4 of the 16
sources at redshift greater than 3, and has resulted in discovery of
two new X-ray jets among the 12 quasars observed so far. For each
individual source in Table~\ref{tab1} the upper limit may allow that
the predicted counts are correct. However, the ensemble of 14 sources
predicts 87 counts, where we have no more than $\approx$40, so that
clearly the average emission is less than our scaling model.

Strictly speaking, we cannot claim that we have discovered ``jets''
according to a common definition \citep{Bridle84} requiring the length
to be at least four times the width for a radio jet. We do have
definite statistical detection of X-ray emission extended outside the
quasar X-ray/radio core.  The emission is statistically connected to a
direction defined \textit{a priori} by radio observations, i.e.,
extremely unlikely to be a background or foreground source not
associated with the system.  However, we have no information on the
structure of the X-ray jet, e.g., whether it is continuous or even
just a single knot of emission.

\begin{table}[h]%
\caption{Comparison of the X-ray dominated jets.\label{tab2}}
\tabcolsep=0pt%
\begin{tabular*}{20pc}{@{\extracolsep\fill}lcll@{\extracolsep\fill}}
\toprule

\textbf{} & \textbf{J0730}$^{\dagger}$  & \textbf{J1405}    & \textbf{J1610}   \\
\midrule
Redshift  & 2.50  &3.209   & 3.118    \\
Live time (ks)  & 19.   & 9.6   & 9.1   \\
Counts  & 38   & 7   & 8    \\
Flux 10$^{-14}$ erg cm$^{-2}$ s$^{-1}$   & 2.7   & 0.9   & 1.      \\
Length (arcsec)  & 12   & 3.5   & 4.6    \\
Surface brightness  & 0.9  & 1.  &0.9  \\ 
\bottomrule
\end{tabular*}
\begin{tablenotes}%%
\item $^{\dagger}$J0730 live time, counts, and length from \cite{Simionescu16}
\end{tablenotes}
\end{table}

\cite{Simionescu16} made the critical discovery of an X-ray jet from
the quasar J0730+4049, that was not coincident with a radio jet.  We
have now discovered two more cases of X-ray jets which are not
coincident with underlying radio jets, establishing that such a class
of object exists. Table~\ref{tab2} compares the observed properties of
these three objects. The flux is taken in the 0.5 to 7 keV band. We
convert to surface brightness in units of 10$^{-14}$ erg cm$^{-2}$
s$^{-1}$ arcsec$^{-2}$ by dividing by the tabulated length of each
jet, and by an assumed 0\farcs25 width for each jet. That width is
roughly 2 kpc at redshifts of a few. The resulting surface brightness
is remarkably similar for the three sources, but with up to a factor
of two uncertainty for the two new jets reported here. This could
indicate similar physical conditions in the three jets, or could just
reflect that they are all near the threshold of detection for a 10 ks
observation. However, a very significant difference is that the jet in
J0730 is detected with an X-ray flux 23\% that of the quasar, while
the two new jets here have only about 2\% of the quasar flux,
consistent with the observations from the survey at lower redshift
\citep{Marshall18}.

These objects are potentially extremely important,
because they can be detected in X-rays at whatever large redshift they
exist, due to their approximately constant surface
brightness. However, discovery of relativistic X-ray jets is difficult,
because only the few percent of radio quasars which are beamed in our
direction are candidate systems, so that it is expensive in observing
time to use pointed \textit{Chandra}  observations for this
purpose. We look forward to the all sky survey by the eROSITA
instrument on the Spectrum X-Gamma satellite \citep{Predehl11} to
provide target systems. While eROSITA does not have the angular
resolution to resolve jets, it should result in unidentified blank
sky sources which are candidates to be very distant ``orphan'' X-ray
jets. 
These candidates will be reasonable targets for \textit{Chandra}
observations, and for the much more powerful  \textit{Lynx} observatory
which is proposed to be the \textit{Chandra} successor, with 30 times
the effective area  and similar angular resolution in the 0.1 to 10 keV
region. Furthermore, \textit{Lynx} with have 800 times larger grasp
than \textit{Chandra} and can therefore do surveys of tens of square
degrees with similar sensitivity. 

\section*{Acknowledgments}
This research was funded by  NASA grant GO8-19077X (\textit{Chandra}) and by
NASA contract NAS8-03060 to the Chandra X-ray Center. 

\bibliography{schwartzRef}%

\begin{thebibliography}{}

\bibitem [\protect \citeauthoryear {%
{Becker}%
, {White}%
\BCBL {}\ \BBA {} {Helfand}%
}{%
{Becker}%
\ \protect \BOthers {.}}{%
{\protect \APACyear {1995}}%
}]{%
Becker95}
\APACinsertmetastar {%
Becker95}%
\begin{APACrefauthors}%
{Becker}, R\BPBI H.%
, {White}, R\BPBI L.%
\BCBL {}\ \BBA {} {Helfand}, D\BPBI J.%
\end{APACrefauthors}%
\unskip\
\newblock
\APACrefYearMonthDay{1995}{}{},
\newblock
\unskip
\newblock
\APACjournalVolNumPages{\apj}{450}{}{559}.
\PrintBackRefs{\CurrentBib}

\bibitem [\protect \citeauthoryear {%
{Bourda}%
, {Collioud}%
, {Charlot}%
, {Porcas}%
\BCBL {}\ \BBA {} {Garrington}%
}{%
{Bourda}%
\ \protect \BOthers {.}}{%
{\protect \APACyear {2011}}%
}]{%
Bourda11}
\APACinsertmetastar {%
Bourda11}%
\begin{APACrefauthors}%
{Bourda}, G.%
, {Collioud}, A.%
, {Charlot}, P.%
, {Porcas}, R.%
\BCBL {}\ \BBA {} {Garrington}, S.%
\end{APACrefauthors}%
\unskip\
\newblock
\APACrefYearMonthDay{2011}{}{},
\newblock
\unskip
\newblock
\APACjournalVolNumPages{\aap}{526}{}{A102}.
\PrintBackRefs{\CurrentBib}

\bibitem [\protect \citeauthoryear {%
{Breiding}%
\ \protect \BOthers {.}}{%
{Breiding}%
\ \protect \BOthers {.}}{%
{\protect \APACyear {2017}}%
}]{%
Breiding17}
\APACinsertmetastar {%
Breiding17}%
\begin{APACrefauthors}%
{Breiding}, P.%
, {Meyer}, E\BPBI T.%
, {Georganopoulos}, M.%
, {Keenan}, M\BPBI E.%
, {DeNigris}, N\BPBI S.%
\BCBL {}\ \BBA {} {Hewitt}, J.%
\end{APACrefauthors}%
\unskip\
\newblock
\APACrefYearMonthDay{2017}{}{},
\newblock
\unskip
\newblock
\APACjournalVolNumPages{\apj}{849}{}{95}.
\PrintBackRefs{\CurrentBib}

\bibitem [\protect \citeauthoryear {%
{Bridle}%
\ \BBA {} {Perley}%
}{%
{Bridle}%
\ \BBA {} {Perley}%
}{%
{\protect \APACyear {1984}}%
}]{%
Bridle84}
\APACinsertmetastar {%
Bridle84}%
\begin{APACrefauthors}%
{Bridle}, A\BPBI H.%
\BCBT {}\ \BBA {} {Perley}, R\BPBI A.%
\end{APACrefauthors}%
\unskip\
\newblock
\APACrefYearMonthDay{1984}{}{},
\newblock
\unskip
\newblock
\APACjournalVolNumPages{\araa}{22}{}{319-358}.
\PrintBackRefs{\CurrentBib}

\bibitem [\protect \citeauthoryear {%
{Cheung}%
\ \protect \BOthers {.}}{%
{Cheung}%
\ \protect \BOthers {.}}{%
{\protect \APACyear {2012}}%
}]{%
Cheung12}
\APACinsertmetastar {%
Cheung12}%
\begin{APACrefauthors}%
{Cheung}, C\BPBI C.%
, {Stawarz}, {\L}.%
, {Siemiginowska}, A.%
, {Gobeille}, D.%
, {Wardle}, J\BPBI F\BPBI C.%
, {Harris}, D\BPBI E.%
\BCBL {}\ \BBA {} {Schwartz}, D\BPBI A.%
\end{APACrefauthors}%
\unskip\
\newblock
\APACrefYearMonthDay{2012}{}{},
\newblock
\unskip
\newblock
\APACjournalVolNumPages{\apjl}{756}{}{L20}.
\PrintBackRefs{\CurrentBib}

\bibitem [\protect \citeauthoryear {%
{Gobeille}%
, {Wardle}%
\BCBL {}\ \BBA {} {Cheung}%
}{%
{Gobeille}%
\ \protect \BOthers {.}}{%
{\protect \APACyear {2014}}%
}]{%
Gobeille14}
\APACinsertmetastar {%
Gobeille14}%
\begin{APACrefauthors}%
{Gobeille}, D\BPBI B.%
, {Wardle}, J\BPBI F\BPBI C.%
\BCBL {}\ \BBA {} {Cheung}, C\BPBI C.%
\end{APACrefauthors}%
\unskip\
\newblock
\APACrefYearMonthDay{2014}{}{},
\newblock
\unskip
\newblock
\APACjournalVolNumPages{ArXiv e-prints}{}{}{},
\newblock
\APAChowpublished {\url{https://arxiv.org/abs/1406.4797}}.
\PrintBackRefs{\CurrentBib}

\bibitem [\protect \citeauthoryear {%
{Gregory}%
, {Scott}%
, {Douglas}%
\BCBL {}\ \BBA {} {Condon}%
}{%
{Gregory}%
\ \protect \BOthers {.}}{%
{\protect \APACyear {1996}}%
}]{%
Gregory96}
\APACinsertmetastar {%
Gregory96}%
\begin{APACrefauthors}%
{Gregory}, P\BPBI C.%
, {Scott}, W\BPBI K.%
, {Douglas}, K.%
\BCBL {}\ \BBA {} {Condon}, J\BPBI J.%
\end{APACrefauthors}%
\unskip\
\newblock
\APACrefYearMonthDay{1996}{}{},
\newblock
\unskip
\newblock
\APACjournalVolNumPages{\apjs}{103}{}{427}.
\PrintBackRefs{\CurrentBib}

\bibitem [\protect \citeauthoryear {%
{Harris}%
\ \BBA {} {Krawczynski}%
}{%
{Harris}%
\ \BBA {} {Krawczynski}%
}{%
{\protect \APACyear {2006}}%
}]{%
Harris06}
\APACinsertmetastar {%
Harris06}%
\begin{APACrefauthors}%
{Harris}, D\BPBI E.%
\BCBT {}\ \BBA {} {Krawczynski}, H.%
\end{APACrefauthors}%
\unskip\
\newblock
\APACrefYearMonthDay{2006}{}{},
\newblock
\unskip
\newblock
\APACjournalVolNumPages{\araa}{44}{}{463-506}.
\PrintBackRefs{\CurrentBib}

\bibitem [\protect \citeauthoryear {%
{Marshall}%
\ \protect \BOthers {.}}{%
{Marshall}%
\ \protect \BOthers {.}}{%
{\protect \APACyear {2018}}%
}]{%
Marshall18}
\APACinsertmetastar {%
Marshall18}%
\begin{APACrefauthors}%
{Marshall}, H\BPBI L.%
, {Gelbord}, J\BPBI M.%
, {Worrall}, D\BPBI M.%
\ et al.\end{APACrefauthors}%
\unskip\
\newblock
\APACrefYearMonthDay{2018}{}{},
\newblock
\unskip
\newblock
\APACjournalVolNumPages{\apj}{856}{}{66}.
\PrintBackRefs{\CurrentBib}

\bibitem [\protect \citeauthoryear {%
{Planck Collaboration}%
, {Aghanim}%
, {Akrami}%
\BCBL {}\ \BBA {} {Ashdown}%
}{%
{Planck Collaboration}%
\ \protect \BOthers {.}}{%
{\protect \APACyear {2017}}%
}]{%
Planck17}
\APACinsertmetastar {%
Planck17}%
\begin{APACrefauthors}%
{Planck Collaboration}%
, {Aghanim}, N.%
, {Akrami}, Y.%
\BCBL {}\ \BBA {} {Ashdown}, M\BPBI e\BPBI a.%
\end{APACrefauthors}%
\unskip\
\newblock
\APACrefYearMonthDay{2017}{}{},
\newblock
\unskip
\newblock
\APACjournalVolNumPages{\aap}{607}{}{A95}.
\PrintBackRefs{\CurrentBib}

\bibitem [\protect \citeauthoryear {%
{Predehl}%
}{%
{Predehl}%
}{%
{\protect \APACyear {2011}}%
}]{%
Predehl11}
\APACinsertmetastar {%
Predehl11}%
\begin{APACrefauthors}%
{Predehl}, P.%
\end{APACrefauthors}%
\unskip\
\newblock
\APACrefYearMonthDay{2011}{}{},
\newblock
{\BBOQ}\APACrefatitle {{eROSITA - A new X-ray All-sky Survey}} {{eROSITA - A
  new X-ray All-sky Survey}}.{\BBCQ}
\newblock
\BIn{} J\BHBI U.~{Ness}\ \BBA {} M.~{Ehle}\ (\BEDS), \APACrefbtitle {The X-ray
  Universe 2011} {The X-ray Universe 2011}\ \BPG~023.,
\newblock
\APAChowpublished {\url{http://adsabs.harvard.edu/abs/2011xru..conf...23P}}.
\PrintBackRefs{\CurrentBib}

\bibitem [\protect \citeauthoryear {%
{Schwartz}%
}{%
{Schwartz}%
}{%
{\protect \APACyear {2002}}%
}]{%
Schwartz02}
\APACinsertmetastar {%
Schwartz02}%
\begin{APACrefauthors}%
{Schwartz}, D\BPBI A.%
\end{APACrefauthors}%
\unskip\
\newblock
\APACrefYearMonthDay{2002}{}{},
\newblock
\unskip
\newblock
\APACjournalVolNumPages{ApJL}{569}{}{L23-L26}.
\PrintBackRefs{\CurrentBib}

\bibitem [\protect \citeauthoryear {%
{Schwartz}%
\ \protect \BOthers {.}}{%
{Schwartz}%
\ \protect \BOthers {.}}{%
{\protect \APACyear {2006}}%
}]{%
Schwartz06}
\APACinsertmetastar {%
Schwartz06}%
\begin{APACrefauthors}%
{Schwartz}, D\BPBI A.%
, {Marshall}, H\BPBI L.%
, {Lovell}, J\BPBI E\BPBI J.%
\ et al.\end{APACrefauthors}%
\unskip\
\newblock
\APACrefYearMonthDay{2006}{}{},
\newblock
\unskip
\newblock
\APACjournalVolNumPages{\apj}{640}{}{592-602}.
\PrintBackRefs{\CurrentBib}

\bibitem [\protect \citeauthoryear {%
{Siemiginowska}%
\ \protect \BOthers {.}}{%
{Siemiginowska}%
\ \protect \BOthers {.}}{%
{\protect \APACyear {2003}}%
}]{%
Siemiginowska03}
\APACinsertmetastar {%
Siemiginowska03}%
\begin{APACrefauthors}%
{Siemiginowska}, A.%
, {Smith}, R\BPBI K.%
, {Aldcroft}, T\BPBI L.%
, {Schwartz}, D\BPBI A.%
, {Paerels}, F.%
\BCBL {}\ \BBA {} {Petric}, A\BPBI O.%
\end{APACrefauthors}%
\unskip\
\newblock
\APACrefYearMonthDay{2003}{}{},
\newblock
\unskip
\newblock
\APACjournalVolNumPages{\apjl}{598}{}{L15-L18}.
\PrintBackRefs{\CurrentBib}

\bibitem [\protect \citeauthoryear {%
{Simionescu}%
\ \protect \BOthers {.}}{%
{Simionescu}%
\ \protect \BOthers {.}}{%
{\protect \APACyear {2016}}%
}]{%
Simionescu16}
\APACinsertmetastar {%
Simionescu16}%
\begin{APACrefauthors}%
{Simionescu}, A.%
, {Stawarz}, {\L}.%
, {Ichinohe}, Y.%
\ et al.\end{APACrefauthors}%
\unskip\
\newblock
\APACrefYearMonthDay{2016}{}{},
\newblock
\unskip
\newblock
\APACjournalVolNumPages{\apjl}{816}{}{L15}.
\PrintBackRefs{\CurrentBib}

\bibitem [\protect \citeauthoryear {%
{Worrall}%
\ \BBA {} {Birkinshaw}%
}{%
{Worrall}%
\ \BBA {} {Birkinshaw}%
}{%
{\protect \APACyear {2006}}%
}]{%
Worrall06}
\APACinsertmetastar {%
Worrall06}%
\begin{APACrefauthors}%
{Worrall}, D\BPBI M.%
\BCBT {}\ \BBA {} {Birkinshaw}, M.%
\end{APACrefauthors}%
\unskip\
\newblock
\APACrefYearMonthDay{2006}{}{},
\newblock
{\BBOQ}\APACrefatitle {{Multiwavelength Evidence of the Physical Processes in
  Radio Jets}} {{Multiwavelength Evidence of the Physical Processes in Radio
  Jets}}.{\BBCQ}
\newblock
\BIn{} D.~{Alloin}\ (\BED), \APACrefbtitle {Physics of Active Galactic Nuclei
  at all Scales} {Physics of Active Galactic Nuclei at all Scales}\ \BVOL~693,
  \BPG~39.
\newblock
\begin{APACrefDOI} \doi{10.1007/3-540-34621-X_2} \end{APACrefDOI}
\PrintBackRefs{\CurrentBib}

\bibitem [\protect \citeauthoryear {%
{Yang}%
, {Gurvits}%
, {Lobanov}%
, {Frey}%
\BCBL {}\ \BBA {} {Hong}%
}{%
{Yang}%
\ \protect \BOthers {.}}{%
{\protect \APACyear {2008}}%
}]{%
Yang08}
\APACinsertmetastar {%
Yang08}%
\begin{APACrefauthors}%
{Yang}, J.%
, {Gurvits}, L\BPBI I.%
, {Lobanov}, A\BPBI P.%
, {Frey}, S.%
\BCBL {}\ \BBA {} {Hong}, X\BHBI Y.%
\end{APACrefauthors}%
\unskip\
\newblock
\APACrefYearMonthDay{2008}{}{},
\newblock
\unskip
\newblock
\APACjournalVolNumPages{\aap}{489}{}{517-524}.
\PrintBackRefs{\CurrentBib}

\end{thebibliography}

\end{document}